\documentclass[aps,prl,reprint,nofootinbib,floatfix]{revtex4-2}

\usepackage[utf8]{inputenc}
\usepackage[T1]{fontenc}
\usepackage{amsmath,amssymb,bm,graphicx,microtype}
\usepackage{tikz}
\usetikzlibrary{calc,arrows.meta}
\usepackage[colorlinks=true,citecolor=blue,linkcolor=blue,urlcolor=blue]{hyperref}

\newcommand{\EE}{\mathbb{E}}

\begin{document}

\title{Spectral Signatures of Replica Symmetry Breaking in Optimization-Induced Random Matrices}

\author{Isaac P\'erez Castillo}
\affiliation{Departamento de F\'isica, Universidad Aut\'onoma Metropolitana-Iztapalapa, San Rafael Atlixco 186, Ciudad de M\'exico 09340, Mexico}

\begin{abstract}
We study optimization-induced matrix ensembles generated by Gibbs measures. The same quenched disorder that weights configurations also supplies the matrix entries observed on them. For glassy Gibbs measures this raises a natural question: does the induced spectrum inherit the underlying glassy Gibbs geometry? In a dense tensor optimization model we find a selective answer. A single induced matrix has a universal leading bulk that washes out the glassy organization. The difference of two matrices built from independent thermal samples in the same disorder does not: its spectrum gives an explicit image of the glassy Gibbs geometry, encoded by the distribution of mutual overlaps between samples. Parisi theory and Monte Carlo confirm this mechanism across simple and glassy phases.
\end{abstract}
\maketitle

Random matrix theory provides a statistical language for spectra in high-dimensional systems whose microscopic description is too detailed to be useful at the scale of the observable. Its classical formulation begins by prescribing an ensemble, such as a Wigner or invariant ensemble, and then asking for the limiting spectral law and fluctuation theory \cite{Wigner1955,Mehta2004,AndersonGuionnetZeitouni2010,PasturShcherbina2011}. Much of modern random matrix theory asks how this picture changes when the matrix entries are no longer independent in the simplest sense. Variance profiles, deformations, tensor contractions, sparsity, topology, and conditioning replace the scalar semicircle mechanism by structured Dyson equations, outlier problems, or cavity distributions \cite{AjankiErdosKruger2019,BaikBenArousPeche2005,BenaychGeorgesNadakuditi2011,KnowlesYin2014,GoulartCouilletComon2022}. The common theme is that a spectrum can be more than a universal null statistic: it can be an observable of the mechanism that generated the matrix.

A parallel lesson comes from the statistical-mechanical approach to random matrices. The Edwards--Jones representation~\cite{EdwardsJones1976} rewrites spectral observables as Gaussian partition functions, and replica or cavity methods then convert the spectral problem into self-consistency equations for resolvents or fields. This route has been developed for sparse symmetric and covariance matrices, non-Hermitian sparse matrices, random graphs with topological correlations, diluted Wishart-type ensembles, and eigenvalue-count large deviations \cite{RogersTakedaPerezCastilloKuhn2008,RogersPerezCastillo2009,RogersPerezVicenteTakedaPerezCastillo2010,DupicPerezCastillo2014,MetzPerezCastillo2016,PerezCastilloMetz2018DilutedWishart,PerezCastilloMetz2018,RamosSanchezGuzmanGonzalezPerezCastilloMetz2021,PerezCastillo2022DilutedWishart,PerezCastilloGuzmanGonzalez2025DilutedWishart,PerezCastillo2026StatMechRMT}. More recently, matrices have been constructed from equilibrium configurations of interacting systems, so that their correlations are inherited from a Boltzmann measure rather than postulated entry by entry \cite{SaberiSaberMoessner2024,OnderSaberiMoessner2026,OnderSaberiMoessner2026Overlap}. This suggests a sharper question: if a matrix is generated by a Gibbs measure, which part of the Gibbs-state geometry survives in the spectrum?

This question is especially natural for glassy Gibbs measures. In a high-temperature phase a typical pair of configurations has one characteristic overlap between configurations, and many observables reduce to functions of that scalar. In a broken phase, the overlap itself becomes distributed: one-step breaking produces separated inter-state and intra-state overlaps, while full breaking produces a continuous hierarchy. The order parameter is therefore not the magnetization or the energy of a single sample, but the joint law of this continuous hierarchy of overlaps. It is consequently not automatic that spectra of matrices generated by a glassy Gibbs measure retain this structure; the question is which spectral observable, if any, is sensitive to it.

To answer this question more concretely, we focus here on glassy Gibbs measures introduced in optimization-induced matrix ensembles. A Gaussian tensor first defines an energy landscape on Boolean configurations and hence a Gibbs measure. A configuration sampled from that measure is then used to contract the same tensor into a matrix. The matrix law is therefore not specified independently of the landscape; it is generated by it. This construction differs from an ordinary tensor-contraction ensemble because the contraction direction is correlated with the tensor being contracted. It also differs from recent Boltzmann-generated matrix ensembles, where spectra inherit equilibrium correlations of sampled configurations \cite{SaberiSaberMoessner2024,OnderSaberiMoessner2026,OnderSaberiMoessner2026Overlap}. Here the inherited structure is the glassy Gibbs geometry of a random optimization problem: the same disorder organizes the Gibbs measure and then supplies the matrix entries observed along its thermal samples. The observable is therefore not the free energy or the overlap law alone, but the spectrum of the induced matrix ensemble.

The concrete model is the dense Boolean $p$-body maximum-average subtensor problem ($p$-MAS), a tensor generalization of the maximum-average submatrix problem studied in statistics, random optimization, and spin-glass theory \cite{MadeiraOliveira2004,ShabalinEtAl2009,SunNobel2013,BhamidiDeyNobel2017,GamarnikLi2018}. In the submatrix problem one searches for a row--column subset of prescribed relative size $m$ with anomalously large average value; in the tensor problem one searches for a selected set, again occupying a fraction $m$ of the indices, whose induced $p$-body average is large. The Gibbs formulation replaces the hard optimization problem by a finite-temperature ensemble over candidate subsets. The dense submatrix problem admits a constrained mean-field spin-glass formulation with nontrivial replica-symmetric, one-step replica-symmetry-breaking (1RSB), and full replica-symmetry-breaking (FRSB) regimes \cite{ErbaKrzakalaPerezOrtizZdeborova2024}. The subtensor extension preserves the same conceptual structure for $p$-body interactions \cite{ErbaKupferschmidPerezOrtizZdeborova2026}, while rigorous work on large-average subtensors has clarified related ground-state and overlap phenomena \cite{HegadeKizildag2025}. This makes dense $p$-MAS a useful testbed: its Gibbs measure has a Parisi-type overlap structure, and the same disorder tensor has a canonical contraction into matrices. The aim is not to improve the optimization algorithm or to locate the maximum-average subtensor threshold. It is to ask what spectral information remains when the matrix itself is produced by the optimization Gibbs measure. This distinction is important because a matrix generated from an optimized or thermally selected configuration is not a generic tensor contraction: the contraction direction has been chosen by the same disorder that supplies the matrix entries.

Our central result is that this transfer of information from Gibbs measure to spectrum is selective. A single induced matrix has a universal dense bulk at leading order: once the density of selected Boolean variables is fixed, the typical spectral density depends on $p$ and $m$, but not on the detailed organization of the Gibbs measure. Thus the most immediate spectral observable is blind to the glassy Gibbs geometry. The correct observable is instead two-replica. One samples two thermal replicas in the same disorder, constructs two induced matrices, and studies, for instance, their difference. At fixed overlap this difference matrix has an explicit block-resolvent law, and averaging over the Gibbs overlap distribution produces a spectral transform of the Parisi order parameter. In this way the overlap geometry of the optimization landscape becomes visible as a random-matrix density.

The mechanism is already visible even before averaging over the Gibbs measure. Conditional on one selected set, the contraction produces a single active block whose leading covariance is fixed by the set size; the normalized bulk therefore cannot distinguish how the Gibbs measure organizes its typical configurations. Conditional on two selected sets in the same disorder, the two contractions share tensor entries on their common support. Their covariance is controlled by the mutual overlap of the two Gibbs samples, so the spectral problem becomes a fixed-overlap block variance-profile problem, followed by averaging over the Gibbs overlap distribution. This is where the glassy order enters the random-matrix calculation.

Indeed, let $J_A$, $|A|=p$, be independent standard Gaussian tensor entries indexed by unordered $p$-tuples. For $\binom{N}{p}$ possible hyperedges and Boolean $N$-tuple $\sigma$ with entries $\sigma_i\in\{0,1\}$, the dense $p$-MAS Hamiltonian at density $m$ is
\begin{equation}
H_J(\sigma)=\sqrt{\frac{p!}{N^{p-1}}}\sum_{\substack{A\subset[N]\\ |A|=p}}J_A\prod_{i\in A}\sigma_i\,,
\label{eq:prl_hamiltonian}
\end{equation}
with Gibbs weight proportional to $e^{\beta H_J(\sigma)}$ on $\sum_i\sigma_i=mN$. The positive sign in the exponential reflects that the energy is used as a maximization objective. For two configurations, $q(\sigma,\tau)=N^{-1}\sum_i\sigma_i\tau_i$, and
\begin{equation}
\frac{1}{N}\EE_J\big[H_J(\sigma)H_J(\tau)\big]=q(\sigma,\tau)^p+o(1)\,.
\label{eq:prl_h_cov}
\end{equation}
The covariance kernel of the Gibbs problem is therefore $\xi(q)=q^p$, so the thermodynamic overlap law is described by the usual Parisi order parameter \cite{Parisi1980,MezardParisiVirasoro1987,Talagrand2011,Panchenko2013}.

The induced matrix is obtained by leaving two tensor indices uncontracted. For $i\ne j$,
\begin{equation}
M_{ij}(\sigma)=\sqrt{\frac{p!}{N^{p-1}}}\,\sigma_i\sigma_j\sum_S J_{\{i,j\}\cup S}\prod_{k\in S}\sigma_k\,.
\label{eq:prl_induced_matrix}
\end{equation}
The diagonal is set to zero, $M_{ii}(\sigma)=0$, and the sum runs over subsets $S\subset[N]\setminus\{i,j\}$ of size $p-2$. Conditioned on a fixed configuration of density $m$, active entries are centered Gaussians with variance $a_m/N+o(N^{-1})$, where
\begin{equation}
a_m=p(p-1)m^{p-2}\,.
\label{eq:prl_am}
\end{equation}
After the selected sites are placed first, $M(\sigma)$ has one active block of size $mN$ and an inactive zero block. The asymptotic spectral density is consequently
\begin{equation}
\rho(\lambda)=(1-m)\delta_0+\frac{1}{2\pi a_m}\sqrt{4ma_m-\lambda^2}\mathbf 1_{|\lambda|\leq2\sqrt{ma_m}}\,,
\label{eq:prl_single_density}
\end{equation}
with $\mathbf{1}_{x}$ an indicator function that returns one when $x$ is true. Equation \eqref{eq:prl_single_density} is a one-configuration bulk spectral law. At fixed density $m$, the matrix $M(\sigma)$ is built by summing the tensor variables attached to selected $p$-tuples over the selected pairs contained in those tuples. The limiting density in \eqref{eq:prl_single_density} describes the self-averaging bulk generated by these many centered contributions. Its independence of $q(\sigma,\tau)$ therefore means that the one-replica bulk spectrum depends on the density of the selected support, not on the overlap with another configuration. Note that Gibbs conditioning may also modify the mean structure of the induced matrix in ways that are separate from the bulk properties described above. In particular, spike-like or finite-rank corrections may appear if the conditioning produces a common mean shift across the selected tuples, so that their contributions reinforce along a distinguished direction of the selected support rather than entering only as centered pair fluctuations. Such effects would be visible through isolated eigenvalues or edge-sensitive observables, while the formula above describes the limiting empirical bulk density.

\begin{figure*}[t!]
\resizebox{\textwidth}{!}{
\begin{tikzpicture}[>=Stealth]
\tikzset{
  spectrumpanel/.style={draw=black,line width=0.4pt,fill=white,inner sep=2pt},
  spectrumconnector/.style={ -{Stealth[length=2.0mm,width=1.5mm]},draw=black!65,line width=0.55pt,opacity=0.72,shorten >=1.5pt,shorten <=1.5pt},
  locatorpoint/.style={circle,fill=white,draw=black,line width=0.65pt,minimum size=5.0pt,inner sep=0pt},
  locatorlabel/.style={font=\Large\bfseries,fill=white,draw=black,line width=0.25pt,rounded corners=0.5pt,inner sep=2.8pt}
}

\node[inner sep=0] (phase) at (0,0)
{\includegraphics[width=20.6cm]{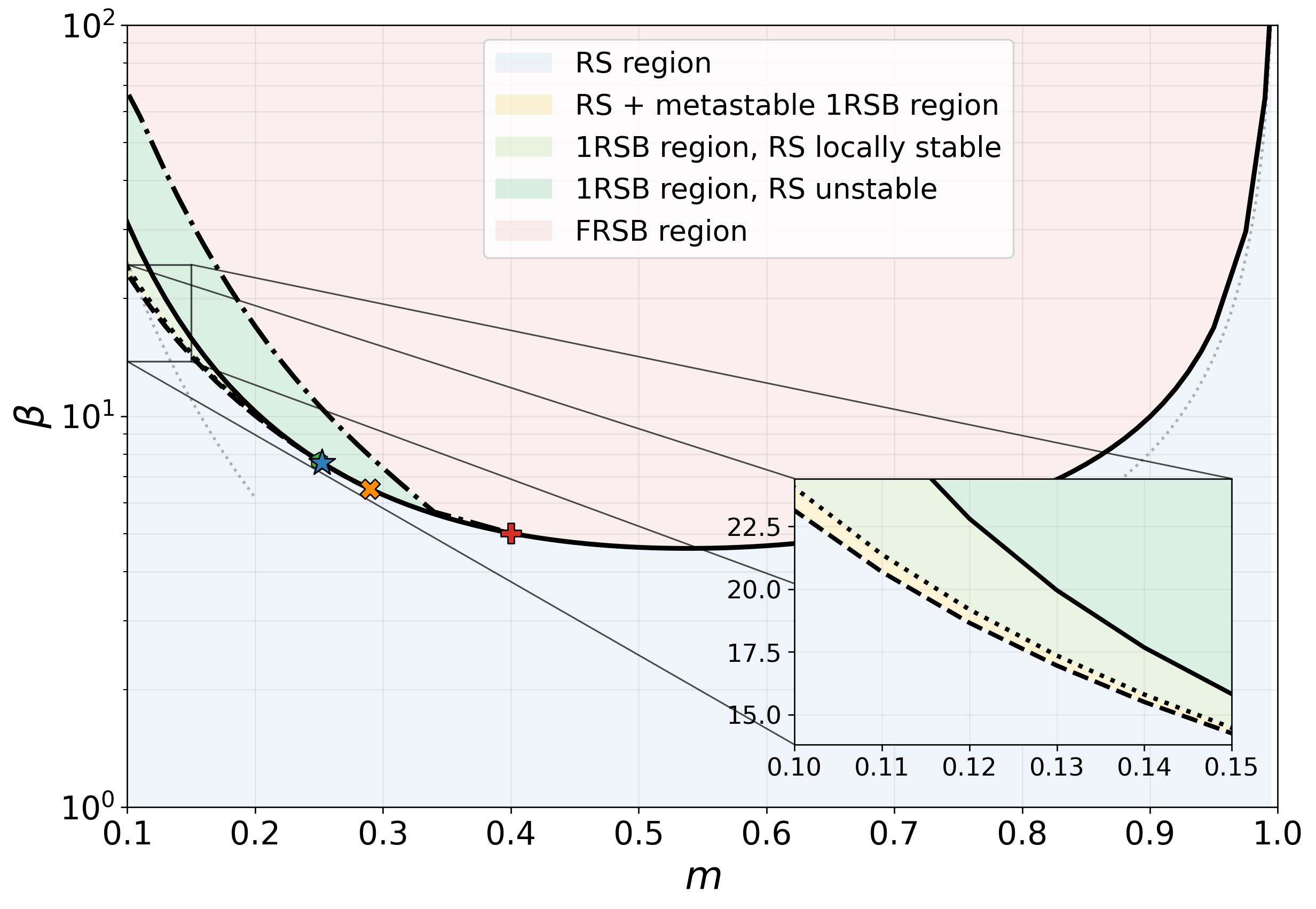}};
\node[spectrumpanel,anchor=south east] (specC) at (-6.7,4.45)
{\includegraphics[width=12.7cm]{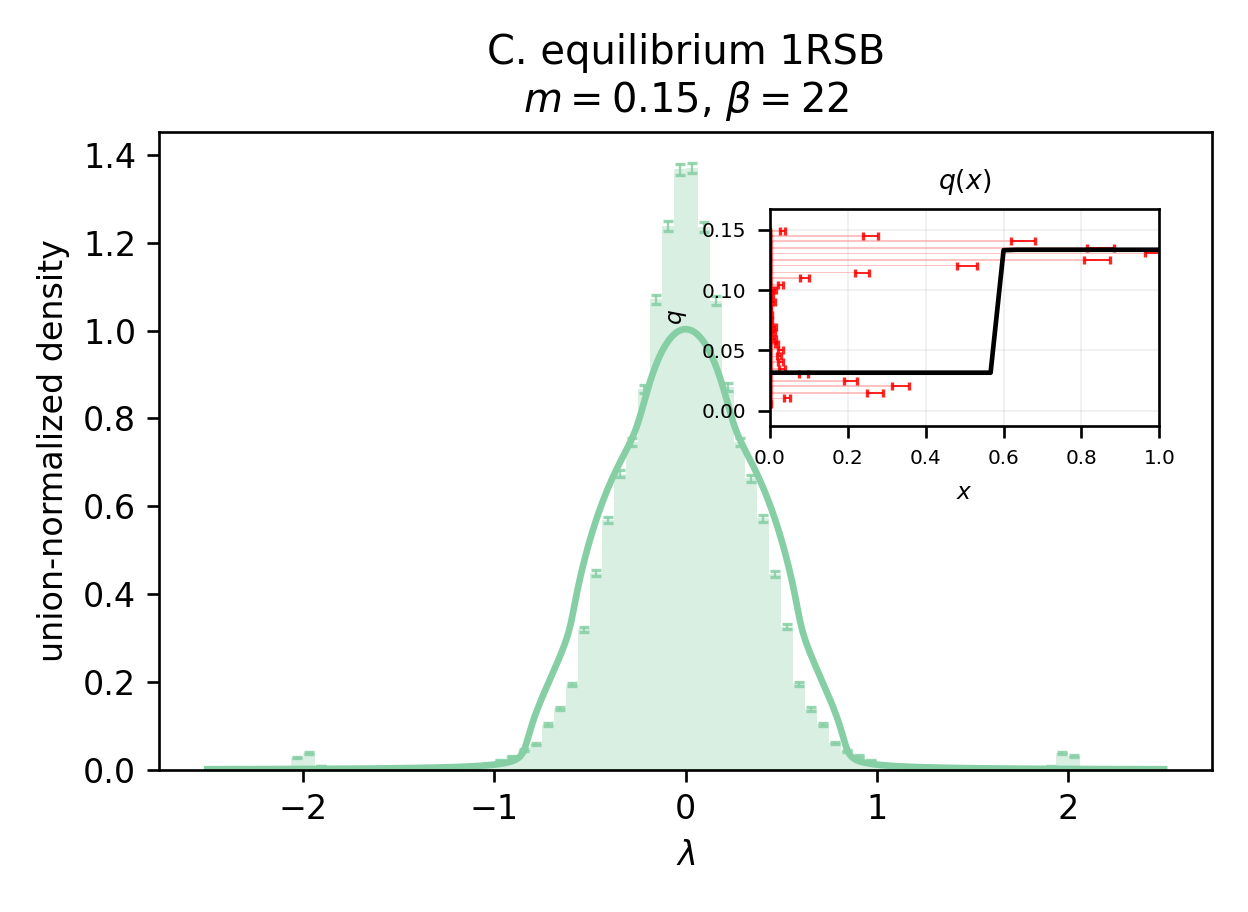}};
\node[spectrumpanel,anchor=south west] (specD) at (6.7,4.45)
{\includegraphics[width=12.7cm]{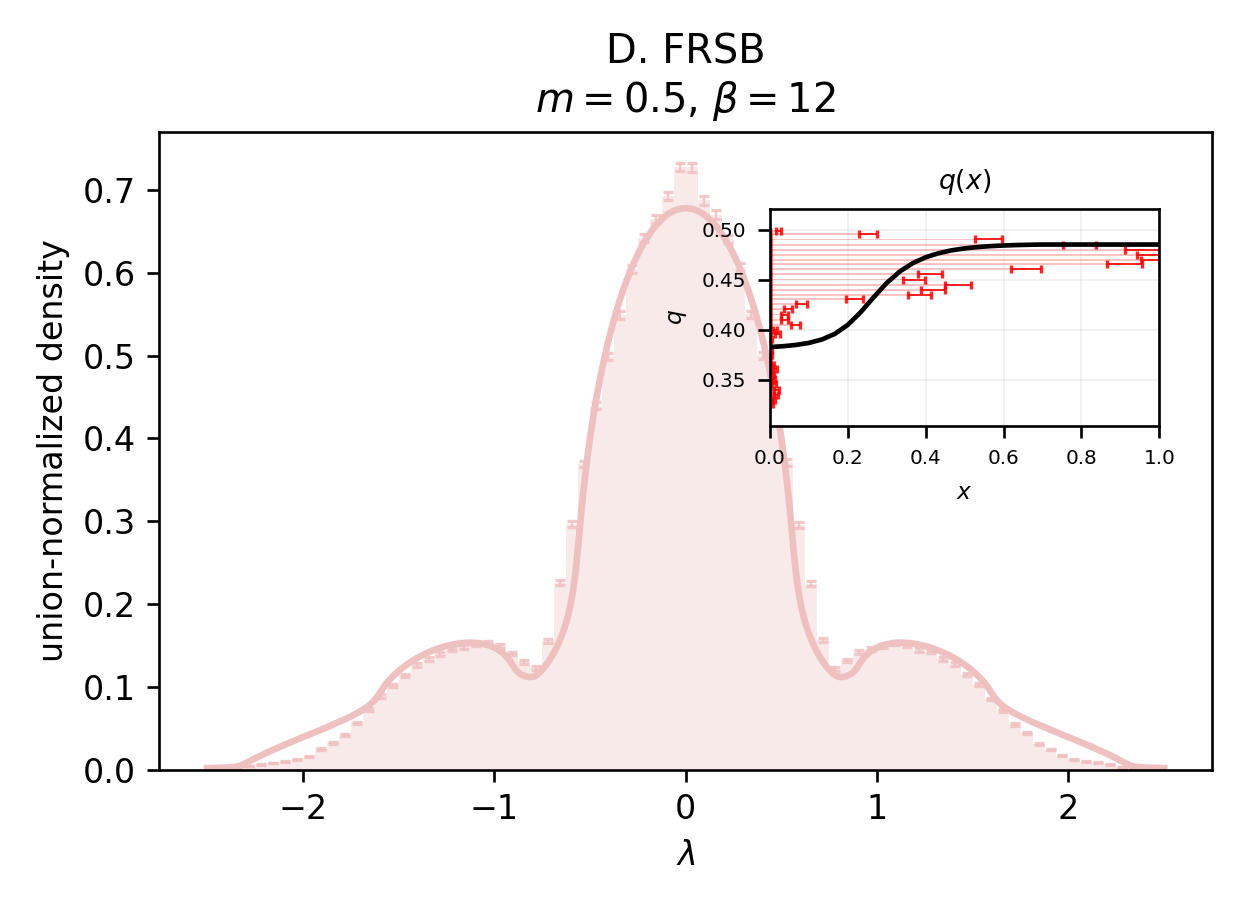}};
\node[spectrumpanel,anchor=north east] (specB) at (-6.7,-4.45)
{\includegraphics[width=12.7cm]{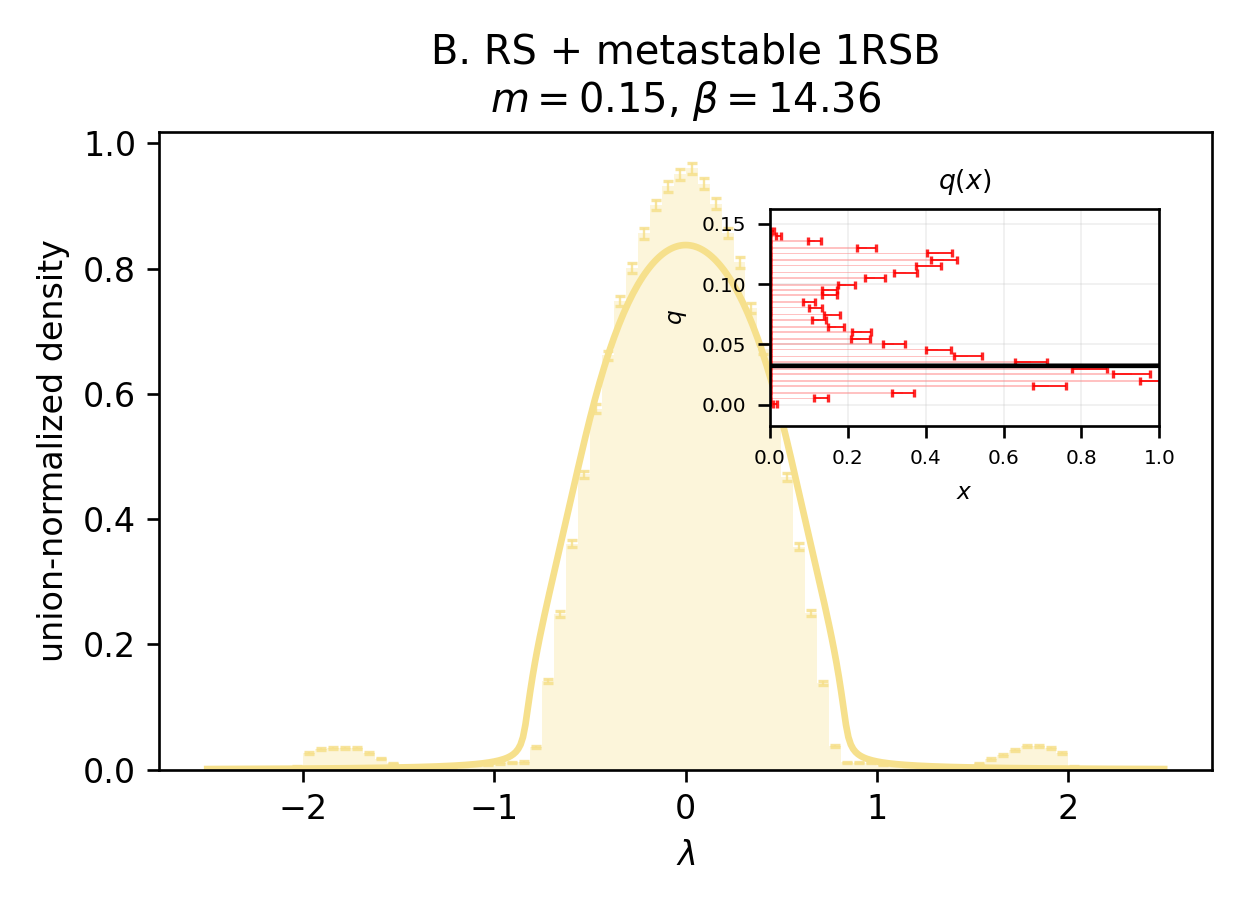}};
\node[spectrumpanel,anchor=north west] (specA) at (6.7,-4.45)
{\includegraphics[width=12.7cm]{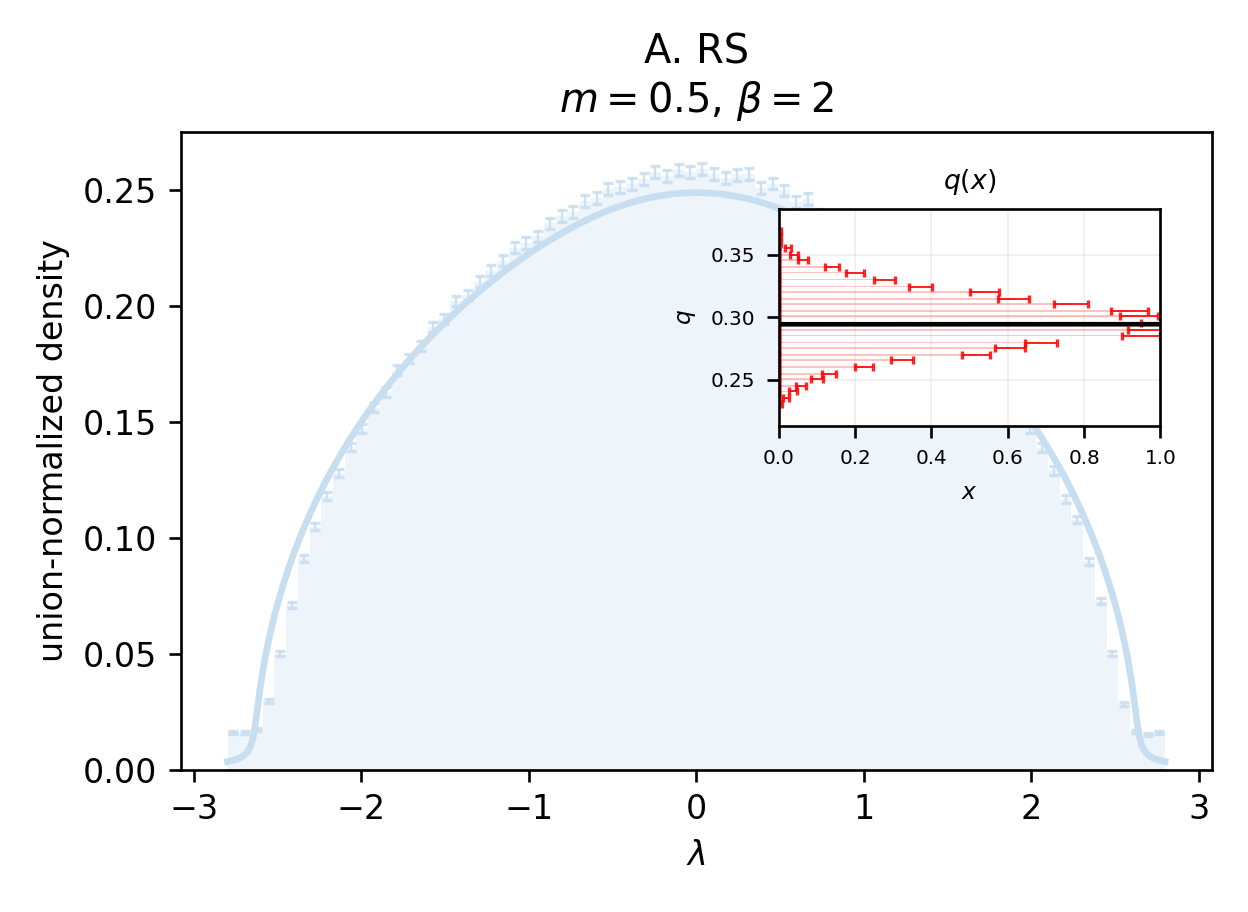}};

\begin{scope}[
  shift={(phase.south west)},
  x={($(phase.south east)-(phase.south west)$)},
  y={($(phase.north west)-(phase.south west)$)}
]
  \coordinate (ptA) at (0.486699,0.239698);
  \coordinate (ptB) at (0.147692,0.607845);
  \coordinate (ptC) at (0.147692,0.687513);
  \coordinate (ptD) at (0.486699,0.574315);
  
  \node[locatorpoint] at (ptA) {};
  \node[locatorpoint] at (ptB) {};
  \node[locatorpoint] at (ptC) {};
  \node[locatorpoint] at (ptD) {};

  \node[locatorlabel,anchor=north west] at ($(ptA)+(0.010,-0.006)$) {A};
  \node[locatorlabel,anchor=north east] at ($(ptB)+(-0.010,-0.006)$) {B};
  \node[locatorlabel,anchor=south east] at ($(ptC)+(-0.010,0.010)$) {C};
  \node[locatorlabel,anchor=west] at ($(ptD)+(0.010,0.010)$) {D};
\end{scope}

\draw[spectrumconnector] (specC.south east) to[out=-18,in=150] (ptC);
\draw[spectrumconnector] (specD.south west) to[out=-160,in=35] (ptD);
\draw[spectrumconnector] (specB.north east) to[out=22,in=-165] (ptB);
\draw[spectrumconnector] (specA.north west) to[out=155,in=-35] (ptA);

\end{tikzpicture}%
}
\caption{Dense Boolean $p$-MAS phase structure and induced two-replica spectra for $p=3$. The central panel shows the phase diagram in the density--inverse-temperature plane, with axes $m$ and $\beta$; the colored legend identifies the filled phase regions. The black transition curves are the replica-symmetric instability (\protect\raisebox{0.35ex}{\textcolor{black}{\rule{1.25em}{0.8pt}}}), the dynamic $x=1$ 1RSB line (\protect\raisebox{0.35ex}{\textcolor{black}{\rule{0.45em}{0.8pt}\hspace{0.14em}\rule{0.45em}{0.8pt}}}), the static $x=1$ 1RSB condensation line (\protect\raisebox{0.35ex}{\textcolor{black}{\rule{0.12em}{0.8pt}\hspace{0.13em}\rule{0.12em}{0.8pt}\hspace{0.13em}\rule{0.12em}{0.8pt}\hspace{0.13em}\rule{0.12em}{0.8pt}}}), and the Gardner instability (\protect\raisebox{0.35ex}{\textcolor{black}{\rule{0.55em}{0.8pt}\hspace{0.12em}\rule{0.12em}{0.8pt}\hspace{0.12em}\rule{0.55em}{0.8pt}}}). The gray dotted curves (\protect\raisebox{0.35ex}{\textcolor[HTML]{777777}{\rule{0.12em}{0.8pt}\hspace{0.13em}\rule{0.12em}{0.8pt}\hspace{0.13em}\rule{0.12em}{0.8pt}\hspace{0.13em}\rule{0.12em}{0.8pt}}}) show the RS asymptotic forms as $m\to0$ and $m\to1$. When present, the blue star (\textcolor[HTML]{2C7FB8}{$\star$}) marks the static--RS endpoint, the orange cross (\textcolor[HTML]{FF8C00}{$\mathbf{X}$}) marks the endpoint of the dynamical $x=1$ 1RSB spinodal, the red filled plus (\protect\raisebox{0.05ex}{\textcolor[HTML]{D73027}{\bfseries +}}) marks the Gardner--RS junction, and the green marker (\textcolor[HTML]{44AA44}{$\blacklozenge$}) marks a dynamic--static crossing. The inset magnifies the selected low-density window; the black rectangle and connectors indicate the magnified region. The surrounding panels show the union-normalized spectral density of $Y_{-1}=M(J,\sigma^1)-M(J,\sigma^2)$ at representative points A--D. Smooth curves are the overlap spectral transform, Eq. \eqref{eq:prl_overlap_transform}; histograms with error bars are finite-temperature Monte Carlo data for $N=200$, obtained from two independent real replicas using the same fixed disorder realization at all four phase points. Insets show the theoretical $q(x)$ and the measured overlap histogram. For each phase point, $4\times10^3$ replica-pair measurements are collected after $10^4$ equilibration sweeps, with consecutive measurements separated by $20$ sweeps.}
\label{fig:phase_spectral_prl}
\end{figure*}
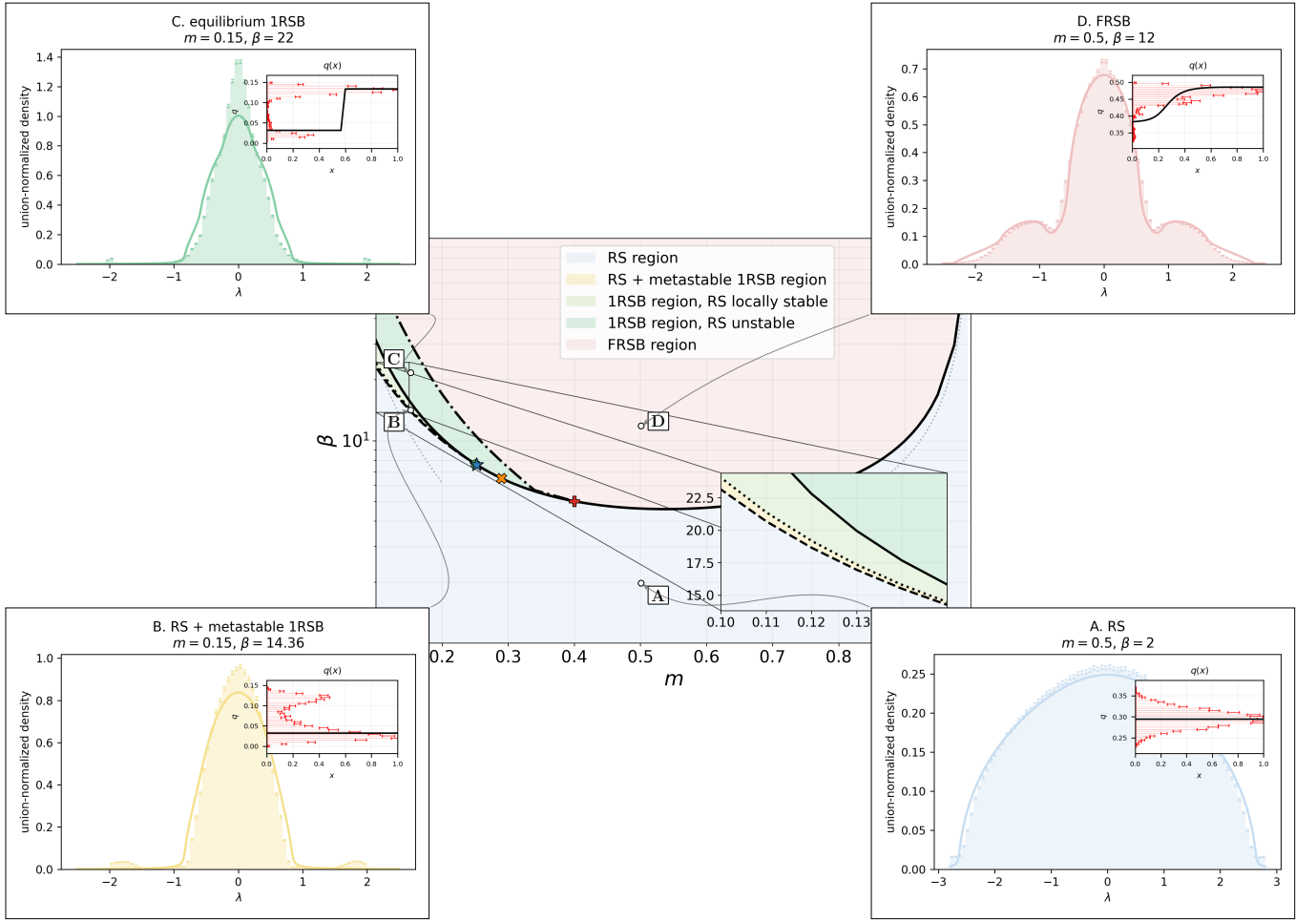

Now take two independent Gibbs samples $\sigma^1,\sigma^2$ at the same disorder, let $M^a=M(J,\sigma^a)$, and consider $Y_t=M^1+tM^2$. At fixed overlap $q=q_{12}$, the indices split into four groups: $A$, selected only by $\sigma^1$; $B$, selected only by $\sigma^2$; $C$, selected by both; and $O$, selected by neither. Their densities are $m-q$, $m-q$, $q$, and $1-2m+q$, respectively. We write $r=m-q$ and denote by $U=A\cup B\cup C$ the union of the two selected supports, whose density is $2m-q$. For $\operatorname{Im}z>0$, set
\begin{equation}
\mathcal G_t(z)=(zI-Y_t)^{-1}\,.    
\end{equation}
The block resolvents are the large-$N$ limits of the average diagonal entries of $\mathcal G_t(z)$ on the three active blocks,
\begin{equation}
g_X^{(t)}(z)=\lim_{N\to\infty}\frac{1}{|X|}\sum_{i\in X}\mathcal G_{t,ii}(z)\,,
\end{equation}
with $X$ labelling the three blocks, $X\in\{A,B,C\}$. For this fixed-overlap block problem, the active-support Stieltjes transform, obtained by averaging diagonal resolvent entries over the union $U$, is
\begin{equation}
\begin{split}
G_t^{\rm union}(z\mid q)&=\lim_{N\to\infty}\frac{1}{|U|}\sum_{i\in U}\mathcal G_{t,ii}(z)\\
&=\frac{r g_A^{(t)}(z)+r g_B^{(t)}(z)+qg_C^{(t)}(z)}{2m-q}\,,
\end{split}
\label{eq:prl_union_transform_fixed_q}
\end{equation}
and the corresponding active-support spectral density, excluding the trivial zero sector on $O$, is
\begin{equation}
\rho_t^{\rm union}(\lambda\mid q)=-\frac{1}{\pi}\lim_{\eta\downarrow0}\operatorname{Im}G_t^{\rm union}(\lambda+i\eta\mid q)\,.
\label{eq:prl_conditional_density}
\end{equation}

It remains to determine the block resolvents. The tensor covariance of matching entries depends on whether the two endpoints lie in one-sided or common blocks. In particular, the common-block covariance of $M^1$ and $M^2$ is controlled by
\begin{equation}
b_q=p(p-1)q^{p-2}=\xi''(q)\,.
\label{eq:prl_bq}
\end{equation}
The appearance of $\xi''$, rather than $\xi$, has a simple origin: the Gibbs Hamiltonian contracts complete $p$-tuples, whereas a matrix entry leaves two tensor indices open. The induced spectral problem therefore probes the two-index marginal covariance of the same disorder that defines the Gibbs landscape.

We now specialize to the difference matrix $Y_{-1}=M^1-M^2$. By symmetry the two one-sided blocks have the same diagonal resolvent, $g_A^{(-1)}=g_B^{(-1)}\equiv g_D$, while $g_C^{(-1)}\equiv g_C$ on the common block. The fixed-overlap equations reduce to
\begin{equation}
\begin{aligned}
g_D(z)&=\frac{1}{z-a_m\{r g_D(z)+q g_C(z)\}}\,,\\
g_C(z)&=\frac{1}{z-2a_m r g_D(z)-2q(a_m-b_q)g_C(z)}\,.
\end{aligned}
\label{eq:prl_difference_resolvent}
\end{equation}
Inserted into the union trace above, these resolvents give the fixed-overlap density $\rho_-^{\rm union}(\lambda\mid q)$ of the difference spectrum.

The dependence on the Gibbs geometry is already visible in, e.g. the conditional second moment,
\begin{equation}
\frac{1}{N}\EE_J\left[\operatorname{Tr}Y_{-1}^2\mid q\right]=2p(p-1)(m^p-q^p)+o(1)\,.
\label{eq:prl_second_moment}
\end{equation}
High-overlap replicas therefore give narrow difference spectra; lower-overlap replicas give broader spectra. If $q=m$, the two configurations coincide at leading density and the difference spectrum collapses to zero.

Let $P_{\beta,m}(dq)$ be the thermodynamic two-replica overlap law of the $p$-MAS Gibbs measure. The observable spectrum is the active-mass-weighted overlap average
\begin{equation}
\rho_{-,\beta}^{\rm union}(\lambda)=\frac{\int (2m-q)\rho_-^{\rm union}(\lambda\mid q)P_{\beta,m}(dq)}{\int(2m-q)P_{\beta,m}(dq)}\,.
\label{eq:prl_overlap_transform}
\end{equation}
The factor $2m-q$ is the density of the union of the two selected supports. It appears both in Eq. \eqref{eq:prl_union_transform_fixed_q} and in Eq.  \eqref{eq:prl_overlap_transform} because the natural normalization of $Y_{-1}$ is on that active union support, rather than on the full set of $N$ indices. This, our central result, is the spectral transform. In a replica-symmetric phase it evaluates a single conditional law. In a 1RSB phase it is a finite mixture associated with inter-state and intra-state overlaps. In a FRSB phase, the measure is represented by a continuous profile $q(x)$ and Eq. \eqref{eq:prl_overlap_transform} becomes the corresponding integral over $x\in[0,1]$. Thus the Parisi order parameter is not inferred indirectly from a scalar statistic, but rather enters an explicit random-matrix map.

The formula also explains why the difference spectrum is the most transparent member of the family $Y_t=M^1+tM^2$. For $t=0$ one recovers the one-matrix law and loses the overlap distribution. For $t=-1$ the common tensor contribution cancels in proportion to the common selected support. The spectrum then narrows continuously as $q$ approaches $m$, and the collapse at $q=m$ is exact at leading order. Other values of $t$ probe reinforcement rather than cancellation, but the same fixed-overlap block law applies.
For $p=3$ we evaluate this construction across the phase diagram displayed in Fig. \ref{fig:phase_spectral_prl}. The transition curves are obtained from the replica thermodynamics of the $p$-MAS and from the stability conditions of the corresponding saddle-point equations: the replica-symmetric replicon, the dynamical one-step saddle-node, the static vanishing-complexity condition, and the Gardner replicon. For each representative point we solve the continuous Full-RSB solution to obtain the profile $q(x)$ and use it to evaluate Eq. \eqref{eq:prl_overlap_transform}. 

The surrounding panels in Fig. \ref{fig:phase_spectral_prl} show the resulting two-replica spectra and their finite-temperature Monte Carlo validation. The RS point produces essentially one conditional spectral law. The metastable 1RSB and equilibrium 1RSB points show the effect of separated overlap values: broadening and changes of shape follow the distribution of inter-state and intra-state overlaps. In the FRSB point the spectrum is a continuous superposition indexed by the Parisi profile. In each panel, the main axes compare the theoretical union-normalized spectral density of $Y_{-1}=M^1-M^2$, obtained by solving the spin-glass problem with replica theory and the saddle-point method and then combining the resulting overlap law with the fixed-overlap block law and Eq.~\eqref{eq:prl_overlap_transform}, against the Monte Carlo estimate obtained by constructing the same matrix difference from two independently sampled replicas at fixed disorder and diagonalizing it. The inset compares the corresponding theoretical Parisi profile $q(x)$ with the overlap histogram measured from the same two-replica Monte Carlo data. The simulations keep the density $m$ fixed by Kawasaki moves and use parallel tempering to equilibrate the system in the rough glassy landscape \cite{MetropolisEtAl1953,Hastings1970,Kawasaki1966,SwendsenWang1986,Geyer1991,HukushimaNemoto1996}. The two real replicas are generated by independent tempering histories for the same disorder sample, so the measured spectra come from the finite-size two-replica Gibbs ensemble at fixed disorder, temperature, and density. The theoretical curves are not fitted to either the spectral histograms or the overlap histograms. The agreement is fairly good across the representative RS, metastable 1RSB, equilibrium 1RSB, and FRSB points, with the observed deviations likely attributed to finite-$N$ effects, sample-to-sample fluctuations, and slower equilibration in the glassy regimes.

The phase diagram explains why the same spectral transform produces qualitatively different spectra in the four representative regimes. In the RS region the overlap law is concentrated on a single sector, and the two-replica spectrum is correspondingly described by essentially one fixed-overlap law. In the low-density 1RSB regime, the dynamical one-step line appears before the static transition, so the phase diagram separates a metastable glassy sector from the equilibrium 1RSB sector. This separation is reflected in the overlap organization and hence in the difference spectrum. In the equilibrium 1RSB region, the spectral density is no longer controlled by the RS overlap alone, but by the mixture of inter-state and intra-state overlap sectors. Beyond the Gardner instability, the one-step description loses stability and the continuous Parisi profile becomes the input to the spectral transform; the FRSB spectrum is therefore a continuous superposition of fixed-overlap laws rather than a finite mixture.

These results identify both a limitation of one-matrix spectra and the advantage of the two-replica observable. A single dense induced matrix is too coarse at the level of its bulk density: after the density of the selected support is fixed, the bulk law does not retain the organization of Gibbs states. The two-replica difference spectrum does retain this information because the two induced matrices are built from the same disorder, and their shared tensor contributions are controlled by the overlap of the two configurations. This makes $Y_{-1}=M^1-M^2$ especially transparent: large overlap produces strong cancellation and a narrower spectrum, whereas small overlap leaves a broader active matrix. The observable therefore converts the replica geometry of the Gibbs measure into a directly measurable spectral effect.

A fuller presentation of the replica derivation of the induced block-resolvent equations, the continuous Full-RSB numerical scheme, the phase-line asymptotics, and the finite-size Monte Carlo analysis will be given elsewhere. For the present Letter, the essential conclusion is already encoded in Eqs. \eqref{eq:prl_single_density} and \eqref{eq:prl_overlap_transform}: optimization-induced matrix ensembles need not reveal the Gibbs geometry at the level of one-matrix spectra, but they can reveal it once the spectral observable is matched to the replica structure. This opens a broader program in which spectra are used as replica-resolved observables of glassy Gibbs measures, in general, and of random optimization problems, in particular. In sparse and diluted versions, where the selected support itself carries nontrivial local structure, topology, localization, and optimization-induced correlations should survive more directly in the bulk.

\bibliographystyle{apsrev4-2}
\bibliography{biblio}

\end{document}